\documentclass[12pt]{academia}
\usepackage{inputenc}
\inputencoding{utf8}
\usepackage{hyperref}
\usepackage{physics}
\usepackage{gensymb}\usepackage[numbers,round]{natbib}
\begin{document}



\articletype{Research article}
\proofstage{Preprint}
\jName{Academia Nano}

\newcommand{\niels}[1]{\textcolor{red}{#1}}

\title{Extended Coupled Cluster approach to Twisted Graphene Layers}


\author{$^{1,2}$Ingvars Vitenburgs, $^{1*}$Niels R. Walet}

\maketitle

\affiliation{$^{1}$Department of Physics and Astronomy, University of Manchester, Manchester M13 9PL, UK. $^{2}$Department of Chemistry, Imperial College London, South Kensington Campus, London, SW7 2AZ, UK.}

\noindent $^{*}$Correspondence:  Niels Walet
\email{$^{*}$niels.walet@manchester.ac.uk}

\section*{Abstract}

A study of correlation effects in twisted bilayer graphene, using the extended coupled cluster method, is presented. This approach considers both self-consistent mean-field and beyond mean-field contributions, and can describe phase transitions in such strongly correlated systems, without further inputs or assumptions. Detailed expressions and a suitable implementation for the method are developed. Combining modern tensor contraction techniques with singular value decomposition, the correlation effects are successfully described in a qualitative manner, including  contributions from the short-range and long-range parts of the Coulomb interaction. The superconducting gap is found to be maximal at a twist angle of $\theta_c = 1.00 \degree$ with a roughly equal combination of s-wave and f-wave components. Using BCS theory, the size of the gap corresponds to a  critical temperature value of $T_\text{c}^\text{BCS} = 0.5$K. This matches qualitatively with experimental data. Within the limitation of the numerical truncations used, a novel candidate for the mechanism behind superconductive phases in twisted bilayer graphene is proposed. 


\section{Introduction}

The mechanisms that give rise to the recently discovered  superconducting and insulating phases in twisted bilayer graphene (TBG) \cite{Cao2018CorrelatedIB, Cao2018UnconventionalS} are of great interest. Theoretical modeling has led to several suggestions for the mechanisms behind these phenomena including long-range electron-electron interactions at Hartree-Fock level, Umklapp scattering and electron-phonon interactions \cite{Cea2021Mechanisms, Lewandowski2021}, but, unfortunately, have yet to yield a fully successful description  of the critical temperature \cite{Cao2018UnconventionalS, Oh2021}. Additionally, there have been many studies \cite{Zhang2020, Xie2020, Bultinck2019, Liu2021, Lee2019, Liu2019, Khalaf2020, Parker2020, Wagner2021} where variation in the symmetries, strain and other details of the realization of TBG lead to a variety of predictions. Nevertheless, previous Hartree-Fock studies \cite{Guinea2018, Cea2019, Cea2021, Pantalen2020, Cea2020, JimenoPozo2023} suggest that electrostatic interactions play a leading role in this system and, hence, electronic correlation effects could be the key to understanding the states. This has also been analysed by other authors using variety of methods \cite{Faulstich2022, Braz2023, Xiao2024}.

The widely used normal coupled-cluster (NCC) method \cite{Bartlett2009} is a standard approach for high precision quantum chemistry, but has some draw-backs for strongly correlated systems, especially when the ground state of the interacting system has a different symmetry than the reference state. These can be partially remedied by combining the NCC method with mean-field methods, thus allowing for a reference state with broken symmetry. An alternative, more direct, approach is to use the extended coupled-cluster (ECC) method. All these coupled-cluster methods share the idea of the use of a reference wave-function acted on by an exponentiated operator consisting of a set of independent (cluster) excitation operators. For both variants the bra state is not the Hermitian conjugate of the ket state, but is instead chosen to ensure unit overlap matrix elements. Whereas in the NCC method this is a linear transformation on the reference states, in the ECC method this is described by yet another exponential. The non-hermiticity, which can be understood through the inherent use of a biorthogonal basis, is a price to pay for the fact that expectation values of operators are finite polynomials in the coefficients of the excitation operators.

The operators in the exponential generating the correlations are usually decomposed in terms of single, double, triple and further excitations relative to the reference state. Here the counter denotes the number of electrons promoted to excited states. The reason we are attracted to the ECC method, which is computationally more expensive than the common NCC approach, is that it incorporates the full Hartree-Fock equations \emph{irrespective of the reference state} and can be used to systematically include additional correlations \cite{Arponen1983, Arponen1987ECC1, Arponen1987ECC2, Arponen1988, Robinson1989, Ligterink01, Snape10, Laestadius2017}. This means that the fully correlated state in the ECC method can actually be orthogonal to the original reference state, which is key to describing phase transitions. In this work, an efficient numerical procedure is developed to implement such calculations for graphene.  

In the case of graphene systems, the most natural choice for the reference state is the non-interacting half-filled Fermi sea, which means that excitations can be labeled by the number of particles and holes relative to this state. This can be most easily done by introducing particle and hole operators to simplify the algebra. It is known \cite{Arponen1983, Arponen1987ECC1, Arponen1987ECC2, Arponen1988} that the ECC singles approach with only one-particle-one-hole excitations is equivalent to the unrestricted Hartree-Fock approximation -- this is not the case for the NCC method, which has also been applied to graphene \cite{Faulstich2022}. Of course, combining NCC with a Hartree-Fock reference state, as is quite common in quantum chemistry \cite{Bartlett2009}, shows some similarity. However, the elegance of the ECC method is that the inherent Hartree-Fock part of the calculation automatically adapts to the many-body correlations, and the extended parametrization allows to describe states that have zero overlap with the reference state, as is the case for phase transitions. As with most other many-body methods the effect of partial filling due to a bias potential applied to graphene can be included by introducing a chemical potential, i.e., by working with the grand potential. 

In the particle and hole operator basis, we introduce a common set of creation operators $\hat{d}_i^\dagger$ that create either a particle or hole, depending on the index $i$. In this form, the singles truncation contains all operators containing the exponential of two creation operators. This then describes both one-particle-one-hole and two-particle and two-hole excitations. Such a combination of correlations operators is fully equivalent to the Hartree-Fock-Bogoliubov formalism without any apriori assumption about the nature of the pairing gap. There is no requirement for the reference state to change from phase to phase, as long as we have a complete enough set of correlation operators.

One of the great advantages of coupled cluster methods over other approaches is that  energy and other expectation values in the reference state is a finite polynomial. These expressions can be written down as tensor contractions, so techniques originally developed for machine learning can be used to efficiently perform the calculations. This corresponds to the search for the extrema of a polynomial energy functional, using tensor manipulations and singular value decomposition (SVD). This means that it is relatively straightforward to efficiently implement such tasks on GPUs. The main goal of this paper is, therefore, to show that the ECC method can be used to describe TBG, find the relevant extremum of the energy functional and describe correlations beyond mean-field truncation, while dealing with pairing and Hartree-Fock correlations on an equal footing. 

The paper is organized as follows: In Section \ref{section:ExtendedCoupledCluster} a very minimal definition of the ECC method and the equations used is given with details relegated to sections \ref{appendix:UndecomposedEquations} and \ref{appendix:DecomposedEquations}. Then, in Section \ref{section:MonolayerGraphene} this approach is validated with monolayer graphene by analyzing the band structure and Fermi level away from half-filling to obtain confidence in the qualitative accuracy of the developed method and its implementation. In Section \ref{section:TwistedBilayerGraphene} superconductivity of TBG near the first magic angle is studied. Further steps are discussed in Section \ref{section:Conclusion}.

\section{Materials and Methods}

\subsection{Extended Coupled Cluster}
\label{section:ExtendedCoupledCluster}

We assume a common set of creation operators $\hat{d}^\dagger_i$, that create all quasi-particle  excitations relative to the uncorrelated reference state $\ket{-}$, and thus $\hat{d}_i \ket{-}=0$ for all $i$. For graphene, the reference state $\ket{-}$ is the half-filled one that describes the non-interacting ground state at charge neutrality. The operators  $\hat{d}^\dagger_i$ create both particles (excitations) and holes (empty states) in $\ket{-}$. The correlation operators contain only even numbers of  operators in order to preserve fermion parity and the indices of the tensors represent the quantum numbers of the system.

A general Hamiltonian can be expressed in second-quantized form in terms of the particle-hole operators as

\begin{equation}
    \hat{H} = \epsilon_{ij} \hat{d}^\dagger_{i} \hat{d}_{j} + \frac{1}{2} V_{ijkl} \hat{d}^\dagger_{i} \hat{d}_{l} \hat{d}^\dagger_{j} \hat{d}_{k} \, .
    \label{equation:Hamiltonian}
\end{equation}

\noindent Here $\epsilon_{ij}$ is a Hermitian matrix and $V_{ijkl}$ is pairwise antisymmetric. In the ECC method \cite{Arponen1983, Arponen1987ECC1, Arponen1987ECC2, Arponen1988} the expectation value of an operator, for example, the Hamiltonian \eqref{equation:Hamiltonian}, is given by

\begin{equation}
    \label{equation:ECC}
    \langle \hat{H} \rangle = \bra{-} e^{\hat{T'}} e^{-\hat{T}} \hat{H} e^{\hat{T}} \ket{-} \, ,
\end{equation}

\noindent where $\hat T$ and $\hat T'$ are two operators with a similar expansion in terms of the creation and annihilation operators $\hat{d}^\dagger_i$ and $\hat{d}_i$ respectively. In the ECC singles and doubles (ECCSD) truncation, the excitation operators $\hat{T}$ and $\hat{T}'$ are fourth order in the single-particle operators and produce

\begin{equation}
\begin{gathered}
    \label{equation:ExcitationOperators}
    \hat{T} = \frac{1}{2!} t_{ij} \hat{d}^\dagger_i \hat{d}^\dagger_j + \frac{1}{4!} t_{ijkl} \hat{d}_{i}^\dagger \hat{d}_{j}^\dagger \hat{d}_{k}^\dagger \hat{d}_{l}^\dagger\, , \\
    \hat{T'} = \frac{1}{2!} t'_{ij} \hat{d}_i \hat{d}_j + \frac{1}{4!} t'_{ijkl} \hat{d}_i \hat{d}_j \hat{d}_k \hat{d}_l \, ,
\end{gathered}
\end{equation}

\noindent where the cluster-amplitude tensors $t$ and $t'$ are fully antisymmetric in their indices. Equation \eqref{equation:ECC} is similar to the more widely used NCC expression, which can be understood as a subset of this method, obtained by expanding $e^{\hat{T'}}$ to first order in $\hat{T'}$. Hence, in standard coupled-cluster terminology Eq. ~\eqref{equation:ExcitationOperators} is the extended coupled cluster method at doubles truncation. The resulting equations for energy and particle number can be found in Sec. ~\ref{appendix:UndecomposedEquations}. These were derived with some help of the Sympy \cite{Sympy} package.

Since the ECC state does not preserve particle number, we need to use the grand potential that contains the chemical potential as a constraint to ensure correct average particle number \cite{Fetter1971}, defined by

\begin{equation}
    \Omega = \frac{\langle \hat{H} - \mu (\hat{N} - N_0) \rangle}{A} \, .
\end{equation}

\noindent Here we have normalized with $A$, which is the area of the unit cell. The particle number constraint has some issues in the numerical calculations, hence, to improve numerical stability and simplify the minimization problem, it is replaced by a physically equivalent, unconstrained quadratic potential following Nogami \cite{Nogami1964} to produce

\begin{equation}
    \label{equation:GrandPotential}
    \Omega_N(\gamma,n_0) = \frac{\langle \hat{H} \rangle}{A} + \gamma \left( \frac{\langle \hat{N} \rangle}{A}  -  n_0 \right)^2 \, .
\end{equation}

\noindent Here $n_0 = \frac{N_0}{A}$ is the electron density per unit cell area and $\gamma$ is a suitably chosen positive parameter. The desired filling can be adjusted by varying $n_0$ and the chemical potential can then be calculated at the unconstrained minimum of $\Omega_N(\gamma,n_0)$ as

\begin{equation}
    \frac{\mu}{A} = - 2 \gamma \left( \frac{\langle \hat{N} \rangle}{A} - n_0 \right) \, .
\end{equation}

Finally, in order to describe single-particle excited states - the band spectra - in a lowest-order approximation, the ECC amplitudes can be kept constant and an additional particle can be added to the ground state by $\ket{-} \rightarrow \hat{d}^\dagger_i \ket{-}$, yielding

\begin{equation}
\begin{gathered}
     \langle \hat{O}_{ij} \rangle = \bra{-} \hat{d}_{\mathbf{k}, i} e^{\hat{T'}} e^{-\hat{T}} \hat{H} e^{\hat{T}} \hat{d}^\dagger_{\mathbf{k}, j} \ket{-} \, , \\ \hat{O} \Psi = \epsilon \Psi \, .
\end{gathered}
\end{equation}

\noindent Here $\epsilon$ are the resulting band energies. This approach is similar to the equation-of-motion framework truncation \cite{Bartlett2009} for the NCC method, where a further set of excitation operators is used to describe the excited state.

Details of all the resulting expressions can be found in Sec.~\ref{appendix:UndecomposedEquations}. These can be implemented straightforwardly for small systems. However, one of the main limitations of the calculations is the amount of computer memory needed to store four-index tensors $t_{ijkl}$, $t'_{ijkl}$, $V_{ijkl}$ and any further intermediate expressions. In order to reduce these rank-4 tensors, appearing in the ECCSD equations, they were decomposed \cite{Hummel2016} using a SVD technique into a sum over a small number of products of rank-2 tensors, thus significantly lowering the computational cost. For a fully antisymmetric tensor such as $t_{ijkl}$ this corresponds to  an expression of the four-index tensors over an antisymmetrized sum over the product of two-index tensors, 

\begin{equation}
\begin{gathered}
    \label{equation:AmplitudesDecomposition}
    t_{ijkl} \approx t_{ij}^a t_{kl}^a  -  t_{ik}^a t_{jl}^a + t_{il}^a t_{kj}^a \, , \\
    t'_{ijkl} \approx  t^{\prime a}_{ij} t^{\prime a}_{kl}  -  t^{\prime a}_{ik} t^{\prime a}_{jl} + t^{\prime a}_{il} t^{\prime a}_{kj} \, .
\end{gathered}
\end{equation}

\noindent We also decompose the potential matrix elements as a sum over pair-like operator, remembering this includes both particle-hole and particle-particle excitations,

\begin{equation}
    V_{ijkl} \approx \overline{V}^a_{ij} V^a_{kl}  \label{equation:PotentialDecomposition}
\end{equation}

\noindent where we determine the dominant components by diagonalizing the Hamiltonian in the basis of two-quasi-particle states for our model space. Here $a$ labels the SVD components, and we retain only those components for the largest eigenvalues. 

Expressing the ECCSD equations in terms of these decompositions, the result at singles level is trivial, since it only involves a change to the potential. At doubles level we require further simplification, and the resulting equations can be found in section \ref{appendix:DecomposedEquations}.

Finally, we apply the automatic differentiation functionality in PyTorch \cite{PyTorch, PyTorchAD} in tandem with parameterizing the tensor contractions using the \textit{einsum} function notation to numerically minimize Eq.~\eqref{equation:GrandPotential}, thus obtaining the ground state energy.

\subsubsection{Extended Coupled Cluster Doubles Equations}
\label{appendix:UndecomposedEquations}

 The energy expectation value for extended coupled cluster at doubles truncation is then found to be

\begin{equation}
\begin{aligned}
     \bra{-} & e^{\hat{T'}} e^{-\hat{T}} \hat{H} e^{\hat{T}} \ket{-}  =  \epsilon_{ik} t_{jk} t'_{ij} + \frac{1}{2} V_{ijkl} t'_{ij} t_{kl}  -  \frac{1}{2} V_{ikkl} t'_{ij} t_{jl} + \frac{1}{2} V_{ijmn} t_{km} t_{ln} (t'_{ij} t'_{kl}  -  2 t'_{ik} t'_{jl} + t'_{ijkl}) \\ & + \frac{1}{12} V_{immn} t_{jkln} t'_{ijkl} + \frac{1}{4} V_{ijmn} t_{klmn} (2 t'_{ik} t'_{jl}  -  t'_{ijkl})  -  \frac{1}{6} \epsilon_{im} t_{jklm} t'_{ijkl} \\ &  + \frac{1}{6} V_{ijop} t_{ko} t_{lmnp} (2 t'_{ik} t'_{jlmn} + 6 t'_{im} t'_{jkln} + 3 t'_{km} t'_{ijln}  -  t'_{ij} t'_{klmn} + 6 t'_{il} t'_{jm} t'_{kn}) \\ & + \frac{1}{72} V_{ijrs} t_{klmr} t_{nops} (6 t'_{ijkl} t'_{mnop} + 9 t'_{ijkn} t'_{lmop}  -  18 t'_{imop} t'_{jkln}  -  2 t'_{iklm} t'_{jnop} \\ & \qquad+ 9 t'_{ij} t'_{kn} t'_{lmop} + 12 t'_{ik} t'_{jm} t'_{lnop} + 18 t'_{ik} t'_{jo} t'_{lmnp} + 72 t'_{ik} t'_{ln} t'_{jmop} \\ & \qquad- 18 t'_{lp} t'_{mn} t'_{ijko} + 6 t'_{ij} t'_{kn} t'_{lp} t'_{mo} + 36 t'_{ik} t'_{jn} t'_{lo} t'_{mp}) \, .
\end{aligned}
\end{equation}

\noindent Here $t$ and $t'$ are defined in Eq. ~\eqref{equation:ExcitationOperators}. The particle number can be expressed as

\begin{equation}
     \langle \hat{N} \rangle = \lambda_{i} \bra{-} e^{\hat{T'}} e^{-\hat{T}} \hat{d}^\dagger_i \hat{d}_i e^{\hat{T}} \ket{-} = \lambda_{i} (-t_{ij} t'_{ij} + \frac{1}{6} t_{ijkl} t'_{ijkl}) \, .
\end{equation}

\noindent Here $\lambda_i$ is $-1$ for hole states and  $+1$ for particle ones. 
The matrix elements of $\hat H$  for single particle  states, used to calculate excited state spectra, is

\begin{equation}
\begin{gathered}
    \bra{-} \hat{d}_x e^{\hat{T'}} e^{-\hat{T}} \hat{H} e^{\hat{T}} \hat{d}^\dagger_y \ket{-} = \epsilon_{xy} + \delta_{xy} (E + V_{xiyi})  -  2 V_{xiym} t_{km} t'_{ik}  -  V_{ijym} t_{xm} t'_{ij} \; + \\ + \; \frac{1}{2} V_{ijyo} t_{xklo} (2 t'_{il} t'_{jk} + t'_{ijkl}) + \frac{1}{3} V_{xiyo} t_{klmo} t'_{iklm} \, ,
\end{gathered}
\end{equation}

\noindent where the states $x$ and $y$ must carry the same momentum. Finally, the expectation value of two annihilation operators, an ingredient for calculating the superconducting gap is

\begin{equation}
    \langle \hat{d}_x \hat{d}_y \rangle = \bra{-} e^{\hat{T'}} e^{-\hat{T}} \hat{d}_x \hat{d}_y e^{\hat{T}} \ket{-} = \frac{1}{8} t_{xijk} t_{ylmn} (t'_{in} t'_{jl} t'_{km}  -  t'_{il}t'_{jkmn}) \, .
\end{equation}

\subsubsection{Decomposed ECCSD Equations}
\label{appendix:DecomposedEquations}

The ECCSD equations in Sec.~\ref{appendix:UndecomposedEquations} can be decomposed using Eq.~\eqref{equation:AmplitudesDecomposition} and Eq.~\eqref{equation:PotentialDecomposition}. Here, the tensors $t$ and $t'$ are antisymmetric in the lower indices, and the index $a$ denotes Einstein summation over a set of component tensors. The rank-4 potential matrix elements $V_{ijkl}$ of the Coulomb force can be naturally decomposed in a sum over density-density interactions \cite{Hummel2016}, i.e., expressed as $\overline{V^a_{il}} V^a_{jk}$, to lower memory consumption.

Even though it is less natural to decompose the potential as a sum over generalized pairing interactions (keep in mind that we work in a particle-hole basis, so this is more than just pairing) as long as we can find an effective procedure, we use the decomposition $V_{ijkl} = \overline{V^a_{ij}} V^a_{kl}$ instead. This simplifies issues with antisymmetrized, which can now be directly incorporated into the tensor $V^a_{ij}$, rather than at the level of the CCM equations. The procedure we choose to determine the dominant components is quite straightforward. We diagonalised the Hamiltonian in the basis of all two-quasi-particle states, making use of momentum conservation to reduce the complexity of the calculations. We then use the largest eigenvalues $\lambda_a$ (in magnitude) and their corresponding eigenvectors $\mathbf{e}_a$ to form the decomposed $V_{ij}^a = \sqrt{\lambda_a} \mathbf{e}^a_{ij}$ potential rank-3 tensor. For simplicity we shall assume the effective range of $a$ in $\tilde{t}^a_{ij}$ and $\tilde{t}^{\prime a}_{ij}$ is the same as that for $\tilde{V}^a_{pq}$. Additionally, we shall also write $V$ for $\tilde V$ and $t$ for $\tilde t$ from here on for simplicity. The difference in the approach to $t$, as opposed to $V$, is that we shall determine the tensors $t^a_{ij}$ and etc. not by an decomposition, but by finding an extremum of the energy functional.

The energy expectation value becomes somewhat more complex,

\begin{equation}
\begin{aligned}
    \bra{-}& e^{\hat{T'}} e^{-\hat{T}} \hat{H} e^{\hat{T}} \ket{-} = \epsilon_{ik} t_{jk} t'_{ij} + \frac{1}{2} \bar{V}^{a}_{ij}V^{a}_{kl} t'_{ij} t_{kl}  -  \frac{1}{2} \bar{V}^{a}_{ik} V^{a}_{kl} t'_{ij} t_{jl} + \frac{1}{2} \bar{V}^{a}_{ij} V^{a}_{mn} t_{km} t_{ln} (t'_{ij} t'_{kl}  -  2 t'_{ik} t'_{jl})
    \\& + \frac{1}{2} t_{km} t_{ln} \overline{V}^{a}_{ij} V^{a}_{mn} t^{\prime b}_{ij} t^{\prime b}_{kl}  -  \frac{1}{12} \overline{V}^{a}_{im} V^{a}_{mn} t^{b}_{jk} t^{b}_{ln} (t^{\prime c}_{ik} t^{\prime c}_{jl} + t^{\prime c}_{il} t^{\prime c}_{jk}) \\&+ \frac{1}{4} \overline{V}^{a}_{ij} V^{a}_{mn} t^{b}_{kl} t^{b}_{mn} (2 t'_{ik} t'_{jl}  -  t^{\prime c}_{ij} t^{\prime c}_{kl})  -  \frac{1}{6} \epsilon_{im} t^{\prime a}_{ij} t^{\prime a}_{kl} (2 t^{b}_{jk} t^{b}_{lm}   -  t^{b}_{jm} t^{b}_{kl}) \\& + \frac{1}{6} \overline{V}^{a}_{ij} V^{a}_{op} t^{b}_{lm} t^{b}_{np} t_{ko} (4 t'_{ik} t^{\prime c}_{jl} t^{\prime c}_{nm}  -  2 t'_{ik} t^{\prime c}_{jn} t^{\prime c}_{lm}  -  12 t'_{im} t^{\prime c}_{jn} t^{\prime c}_{kl}  -  12 t'_{im} t^{\prime c}_{jl} t^{\prime c}_{kn} 
    \\& \qquad\qquad- 6 t'_{im} t^{\prime c}_{jk} t^{\prime c}_{ln}  -  3 t'_{kn} t^{\prime c}_{ij} t^{\prime c}_{lm}  -  2 t'_{ij} t^{\prime c}_{kl} t^{\prime c}_{mn} + t'_{ij} t^{\prime c}_{kn} t^{\prime c}_{lm} + 6 t'_{il} t'_{jm} t'_{kn}) 
    \\& + \frac{1}{72} \overline{V}^{a}_{ij} V^{a}_{rs} t^{b}_{no} t^{b}_{ps} t^{c}_{kl} t^{c}_{mr} 
    \\&\qquad\qquad(12 t^{\prime d}_{ij} t^{\prime d}_{kl} t^{\prime e}_{mn} t^{\prime e}_{op}  -  6 t^{\prime d}_{ij} t^{\prime d}_{kl} t^{\prime e}_{mp} t^{\prime e}_{no} 
    \\& \qquad\qquad+ 36 t^{\prime d}_{ij} t^{\prime d}_{kn} t^{\prime e}_{lm} t^{\prime e}_{op} + 36 t^{\prime d}_{ij} t^{\prime d}_{kn} t^{\prime e}_{lo} t^{\prime e}_{mp} + 9 t^{\prime d}_{ij} t^{\prime d}_{mp} t^{\prime e}_{lk} t^{\prime e}_{on} 
    \\&\qquad\qquad-  72 t^{\prime d}_{im} t^{\prime d}_{jk} t^{\prime e}_{ln} t^{\prime e}_{op} + 72 t^{\prime d}_{ip} t^{\prime d}_{jn} t^{\prime e}_{kl} t^{\prime e}_{mo} + 18 t^{\prime d}_{im} t^{\prime d}_{jl} t^{\prime e}_{kp} t^{\prime e}_{on} 
    \\& \qquad\qquad- 8 t^{\prime d}_{il} t^{\prime d}_{jp} t^{\prime e}_{km} t^{\prime e}_{no}  -  8 t^{\prime d}_{ik} t^{\prime d}_{jn} t^{\prime e}_{lm} t^{\prime e}_{op}  -  2 t^{\prime d}_{im} t^{\prime d}_{jp} t^{\prime e}_{kl} t^{\prime e}_{no}  
    \\& \qquad\qquad+ 36 t'_{ij} t'_{mn} t^{\prime d}_{lk} t^{\prime d}_{op} + 36 t'_{ij} t'_{kp} t^{\prime d}_{lm} t^{\prime d}_{on} + 9 t'_{ij} t'_{mp} t^{\prime d}_{lk} t^{\prime d}_{on} 
    \\ &\qquad\qquad+ 24 t'_{ik} t'_{jl} t^{\prime d}_{mo} t^{\prime d}_{np} + 12 t'_{ik} t'_{jl} t^{\prime d}_{mp} t^{\prime d}_{no} + 36 t'_{ik} t'_{jp} t^{\prime d}_{ml} t^{\prime d}_{no}  -  18 t'_{im} t'_{jp} t^{\prime d}_{lk} t^{\prime d}_{no} 
    \\ &\qquad\qquad+ 144 t'_{ik} t'_{ln} t^{\prime d}_{jm} t^{\prime d}_{op} + 144 t'_{ik} t'_{mn} t^{\prime d}_{jl} t^{\prime d}_{po} + 72 t'_{il} t'_{kp} t^{\prime d}_{jm} t^{\prime d}_{on}  -  72 t'_{ik} t'_{lp} t^{\prime d}_{jm} t^{\prime d}_{on} 
    \\ &\qquad\qquad- 72 t'_{ik} t'_{mp} t^{\prime d}_{jl} t^{\prime d}_{on}  -  12 t'_{ij} t'_{kn} t'_{lp} t'_{mo} + 6 t'_{ij} t'_{mp} t'_{ln} t'_{ko} + 144 t'_{ik} t'_{jp} t'_{mo} t'_{ln} 
    \\ & \qquad\qquad- 72 t'_{ik} t'_{jn} t'_{lp} t'_{mo} + 72 t'_{ik} t'_{jn} t'_{lo} t'_{mp} + 36 t'_{im} t'_{jp} t'_{lo} t'_{kn}) \, .
\end{aligned}
\end{equation}

\noindent Furthermore, the particle number expression is 

\begin{equation}
    \langle \hat{N} \rangle =  \lambda_{i} (-t_{ij} t'_{ij} + \frac{1}{6} t^{p}_{ij} t^{p}_{kl} (3 t^{\prime q}_{ij} t^{\prime q}_{kl} + 2 t^{\prime q}_{il} t^{\prime q}_{kj})) \, .
\end{equation}

\noindent The matrix elements used to calculate excited state energies are defined by

\begin{equation}
\begin{gathered}
    \bra{-} \hat{d}_x e^{\hat{T'}} e^{-\hat{T}} \hat{H} e^{\hat{T}} \hat{d}^\dagger_y \ket{-} = \epsilon_{xy} + \delta_{xy} (E + \overline{V}^{a}_{xi} V^{a}_{yi})  -  2 \overline{V}^{a}_{xi} V^{a}_{ym} t_{km} t'_{ik}  -  \overline{V}^{a}_{ij} V^{a}_{ym} t_{xm} t'_{ij} + \\ + \overline{V}^{a}_{ij} V^{a}_{yo} t^{b}_{xk} t^{b}_{lo} (2 t'_{il} t'_{jk} + t^{\prime c}_{ij} t^{\prime c}_{kl}) + \\ + \frac{1}{2} \overline{V}^{a}_{ij} V^{a}_{yo} t^{b}_{ox} t^{b}_{kl} (2 t'_{il} t'_{jk} + t^{\prime c}_{ij} t^{\prime c}_{kl}) + \\ + \frac{1}{3} \overline{V}^{a}_{xi} V^{a}_{yo} t^{b}_{kl} t^{b}_{mo} (2 t^{\prime c}_{ik} t^{\prime c}_{lm}  -  t^{\prime c}_{im} t^{\prime c}_{kl}) \, .
\end{gathered}
\end{equation}

\noindent Finally, the expectation value required for calculating the superconducting gap is derived to be

\begin{equation}
\begin{gathered}
    \langle \hat{d}_x \hat{d}_y \rangle = \frac{1}{8} t^{a}_{xi} t^{a}_{jk} t^{b}_{yl} t^{b}_{mn} (2 t'_{in} t'_{jl} t'_{km}  -  t'_{kn} t'_{jm} t'_{il}  -  t'_{il} t^{\prime c}_{jk} t^{\prime c}_{mn} + \\ + 4 t'_{im} t^{\prime c}_{jl} t^{\prime c}_{kn} + 4 t'_{il} t^{\prime c}_{jm} t^{\prime c}_{kn}) \, .
\end{gathered}
\end{equation}

\subsection{Code availability}

The codes used to produce the results reported in this paper are available on GitHub \cite{github}.

\section{Results}

\subsection{Monolayer Graphene}
\label{section:MonolayerGraphene}

\begin{figure}
    \centering
    \includegraphics[width=0.5\columnwidth]{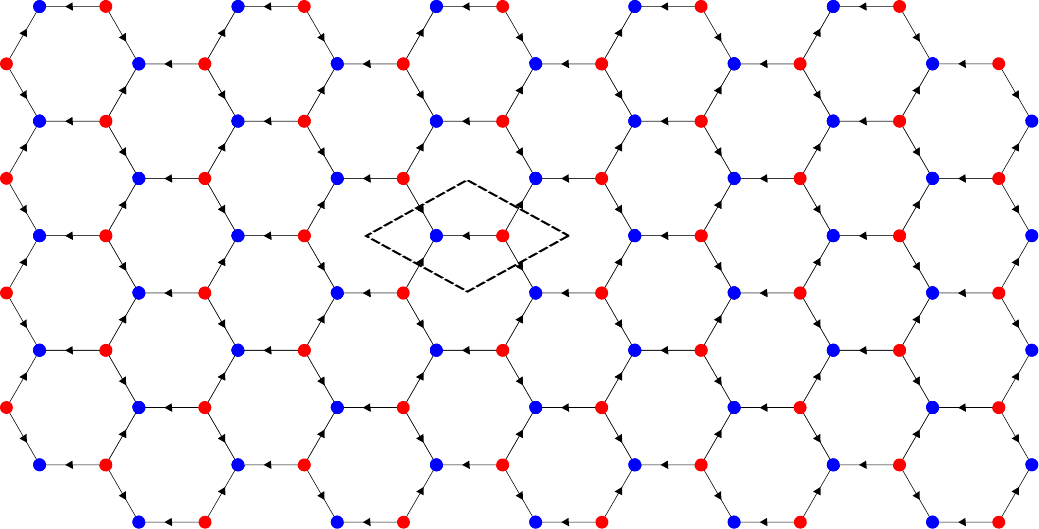}
    \caption{Geometric structure of monolayer graphene. The system is assumed to be described by a nearest-neighbor hopping between the blue and red sites. The dashed region represents the unit cell.}
    \label{figure:MonolayerModel}
\end{figure}

In order to validate our ECCSD implementation, it was tested on a simple model of monolayer graphene \cite{Neto2007}. In the simplest case this is described by a primitive nearest-neighbor hopping model between the blue and red sites, as displayed in Fig. ~\ref{figure:MonolayerModel}.

\noindent This describes the standard semi-metallic spectrum of graphene. A Coulomb interaction between electrons is then added and, due to the use of periodic boundary conditions, the Ewald summation \cite{Ewald1921, Harris1998} technique, in combination with the Widom insertion method \cite{Bakhshandeh2022}, is applied. A simple Monkhorst-Pack grid \cite{Monkhorst1976, Pack1977} was used for sampling the lattice.

\begin{figure}
    \centering
    \includegraphics[width=0.5\columnwidth]{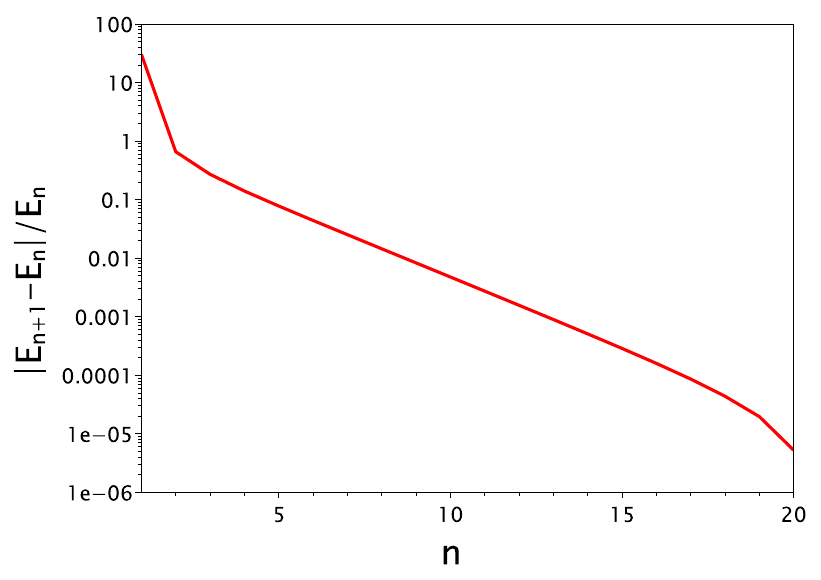}
    \caption{Convergence of the ECCSD simulation energy for monolayer graphene as a step $n$ function of the simulation - a stochastic gradient descent solver. The particle number was observed to converge in a very similar fashion.}
    \label{figure:SimulationConvergence}
\end{figure}

An ECCSD calculation was run for this system, obtaining polynomial convergence, as displayed in Fig. ~\ref{figure:SimulationConvergence}. It was terminated once the relative difference between runs of the stochastic gradient descent solver reached $10^{-6}$, which is a very accurate outcome in single precision, used in order to lower the computational cost.

Generally, to extract asymptotic values of any observable with respect to convergence parameters, extrapolation was performed throughout this study using curves of the form 

\begin{equation}
    \label{equation:ConvergenceCurve}
    O(n) = O_\infty + \frac{A}{B^n} \, .
\end{equation}

\noindent In all further simulations it was made sure that the $n$ for the final convergence step was large enough so that the values of $O(n)$ were within $10 \%$ of the asymptotic value $O_\infty$. This ensures that converged results could be reliably extracted. 

An identical shift of $25 \, $meV, but opposite sign, of the excited state energy was obtained at the Dirac point for both bands at $n_0 = \pm 1$ fillings. A further calculation of the chemical potential yielded a Fermi level of $110 \, $meV. These values are quite close to the experimental ones \cite{Jeon2013} of $\sim 20 \,$meV and $\sim 100 \,$meV, respectively. Furthermore, correlation effects can now be studied by comparing simulations at singles and doubles truncation. Whilst there is negligible impact on the spectrum (changes of the order of one part in a thousand) at a $n_0 = +1$ filling, the Fermi level increased by $40 \, \mathrm{meV}$ when correlation effects were taken into account. Thus, correlations may have quite a significant impact on the electrical conductivity in monolayer graphene.

\subsection{Superconductivity in Twisted Bilayer Graphene}
\label{section:TwistedBilayerGraphene}

\begin{figure}
    \centering
    \includegraphics[width=0.5\columnwidth]{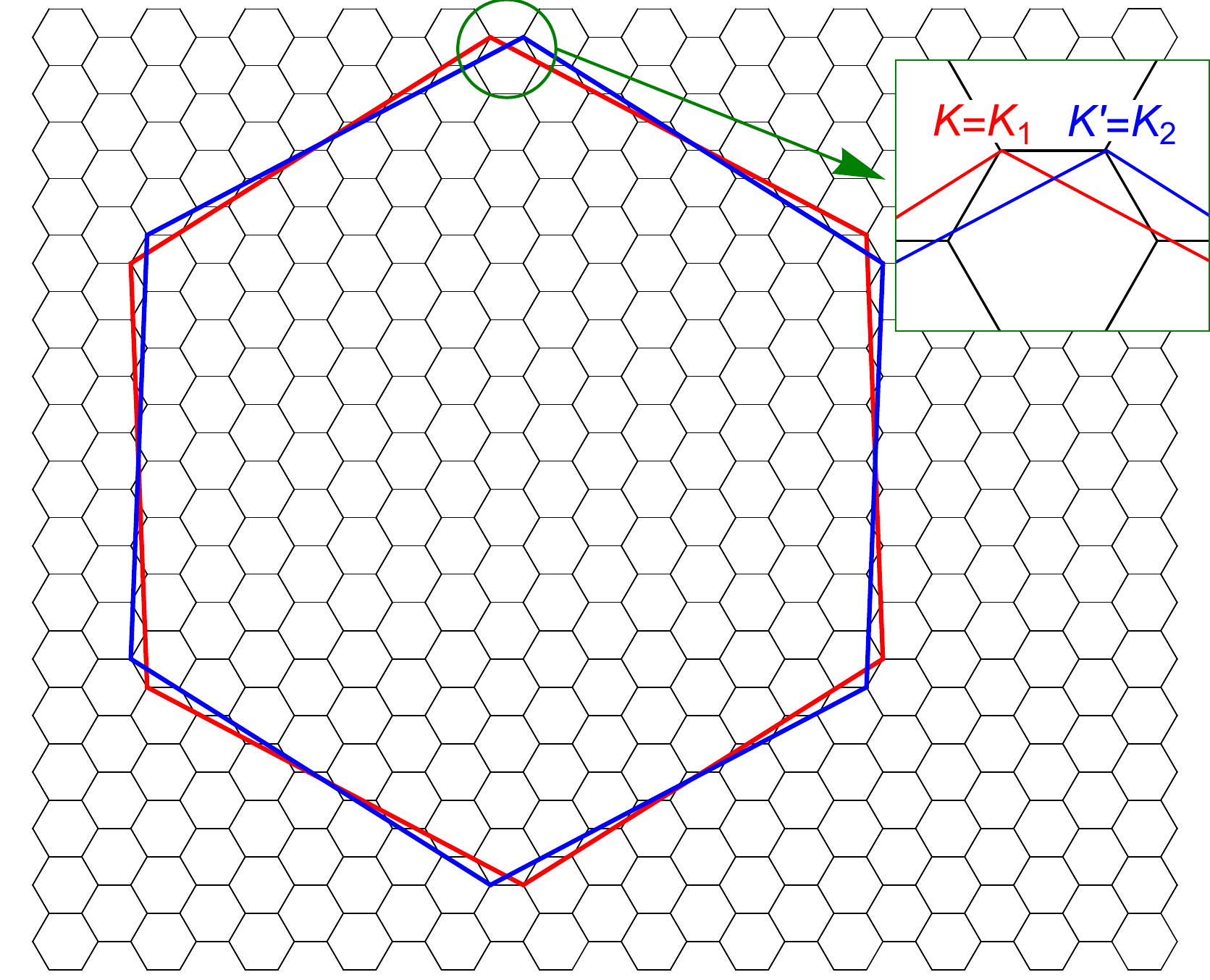}
    \caption{Reciprocal lattice of twisted bilayer graphene. The black hexagons are the TBG unit cells, while the blue and red hexagons are the reciprocal unit cells for each graphene layer. The inset shows that the superlattice $K$ and $K'$ points are the individual $K$ points of the twisted graphene lattices.}
    \label{figure:BilayerModel}
\end{figure}

Having gained confidence in the newly developed method, it is applied to a system of interest - twisted bilayer graphene - to potentially obtain a qualitative description of the reported superconductivity. Even though, in principle, an atomic basis can be used, the dimensionality of the problem becomes extremely large, making it impossible to implement on relatively simple computational resources. Fortunately, TBG is widely modeled by the Bistritzer-MacDonald (BM) model \cite{Bistritzer2010}, which is a low energy continuum description, suitable for small twist angles. The expansion is based on the continuum graphene spectrum at the relevant Dirac points in each layer, as well as a lowest-harmonic expansion of the hopping between the layers, displayed in Fig. ~\ref{figure:BilayerModel}.
 
For a single valley, the BM model is defined by a block Hamiltonian

\begin{equation}
    \hat{H} =  
    \begin{pmatrix}
        H_1 & U \\
        U^\dagger & H_2 \\
    \end{pmatrix}_{ij} \hat{d}_i^\dagger \hat{d}_j \, .
\end{equation}

\noindent The valley number is suppressed in the indices $\{ i \}$ and $\{ j \}$, on a approximation basis, which in turn label the discrete $k$ points, harmonics, cells, bands and spins used. The blocks $H_1$ and $H_2$ represent the continuum graphene Hamiltonian in each of the layers, while $U$ describes the hopping energies defined by the original model values. To include the majority of the contributing processes into this model, the Dirac point, as well as its 6 nearest neighbors (a total of 7 so-called cells) is taken into account \cite{Guinea2018} together with couplings from each one of them to the 3 nearest ones (the so-called harmonics) in the second layer \cite{Guinea2019}. This corresponds to $\sim 80 \%$ of the total short-range interactions. Furthermore, a metallic gate potential \cite{Goodwin2019} is used to model the enclosure of TBG between two layers of hexagonal boron nitride. It is described here by a gate spacing parameter $D = 40$ nm and dielectric permittivity $\epsilon = 10 \epsilon_0$ as

\begin{equation}
    \label{equation:Potential}
    V(\mathbf{q}) = \frac{2 \pi e^2}{\epsilon |\mathbf{a_1} \times \mathbf{a_2}|} \frac{\tanh(D |\mathbf{q}|)}{|\mathbf{q}|}\,.
\end{equation}

\noindent Here the area of the unit cell is $|\mathbf{a_1} \times \mathbf{a_2}| = \frac{\sqrt{3} a^2}{8 \sin(\theta / 2)^2}$ and the same-site energy $U = 17$ meV is used.

The ECCSD simulations were initially run at different truncation levels for a filling of $n_0 = +2$, which has been reported to exhibit potential for insulating to conductive phase transition \cite{Cea2021}. A similar convergence to monolayer graphene was obtained of the stochastic gradient descent solver with around twice as many steps required to reach the relative energy change threshold of $10^{-6}$. Because of the, significantly, smaller first Brillouin zone, in comparison with monolayer graphene, a smaller number of sampling points were expected to be necessary for convergence. This proved to be true, after analyzing Eq. ~\eqref{equation:ConvergenceCurve} type fitted curves, as discussed previously, with only four points required for a qualitative description.

\begin{figure}
    \centering
    \includegraphics[width=0.5\textwidth]{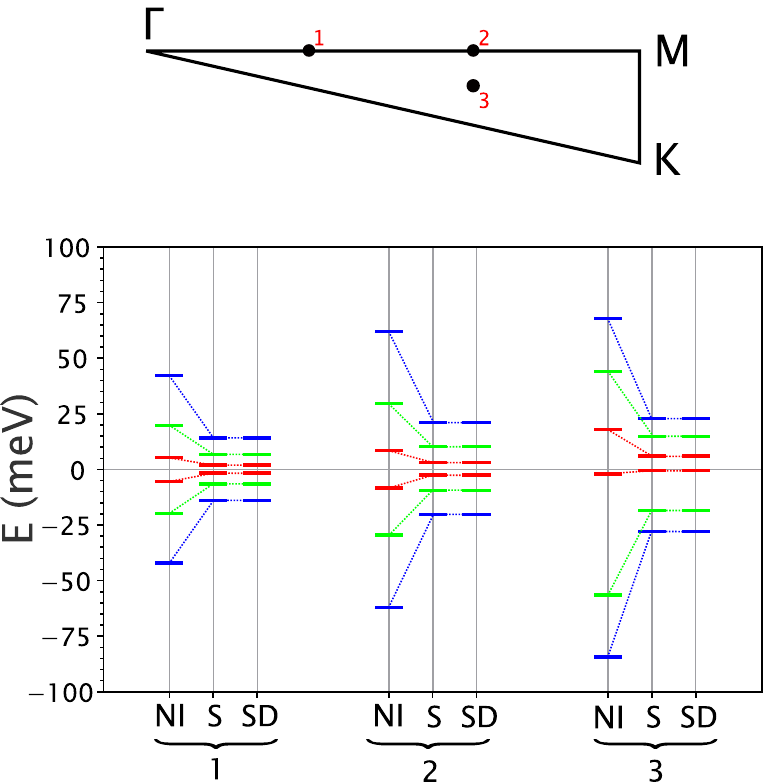}
    \caption{The band structure of twisted bilayer graphene, sampled at three points - two points equidistant on the $K-M$ line and one in the middle of the $\Gamma-K-M$ zone, as shown at the top of the figure. For each point, the non-interacting case and the ECCS and ECCSD coupled cluster truncations are shown.}
    \label{figure:BilayerBands}
\end{figure}

Due to computational cost, only three points were sampled via the Monkhorst-Pack grid at this time - two equidistant ones on the $K-M$ momentum line and one in the middle of the $\Gamma-K-M$ zone. Nevertheless, as displayed in Fig. ~\ref{figure:BilayerBands}, this gave a qualitative description of the mean-field effects. The conclusion that electrostatic interactions have a leading role in the band structure, compressing the band gaps to a few meV, matches previous work \cite{Cea2019, Cea2021}. Furthermore, the effect of correlations was negligible, with a relative contribution of $\sim 10^{-2}$ to the band structure, which is expected to be unchanged by including additional sampling points. A further ECCS calculation of the chemical potential yielded a Fermi level of $25 \, $meV, matching values of $\sim 20 \,$meV from previous studies and signifying occupation of these bands.

In order to make any further calculations feasible, we implement the SVD procedure of Sec. ~\ref{appendix:DecomposedEquations} and restrict the summation range of the third - $a$ - index. In this way the majority of the contributing terms can be included - more specifically, the 16 largest singular values and their associated eigenvectors of the potential $V^{a}_{pq}$ were used, corresponding to $\sim 70 \%$ of the long-range interaction strength. For simplicity the same cut-off was used for $t^{a}_{ij}$. This leads to a significant performance boost because the simulations now fit onto a single NVIDIA 80GB A100 graphics processing unit, designed for fast tensor calculations. Nevertheless, the calculations are limited by the amount of RAM in such systems.

We can calculate \cite{Annett2004} the superconducting gap and the superconducting order parameter via

\begin{equation}
    \Delta_\mathbf{k} =  \sum_{\mathbf{q}} V_{\mathbf{k}, \mathbf{q}} \langle \hat{d}_{\mathbf{q}, \uparrow} \hat{d}_{-\mathbf{q}, \downarrow} \rangle \, .
\end{equation}

\noindent The tensor contractions for this expectation value can be found in section \ref{appendix:UndecomposedEquations}. Implementing this observable for measurements requires an introduction of spin and momenta with opposite signs, totaling $\sim 1000$ sites for a qualitative description with four sampled points.

\begin{figure}
    \centering
    \includegraphics[width=0.5\textwidth]{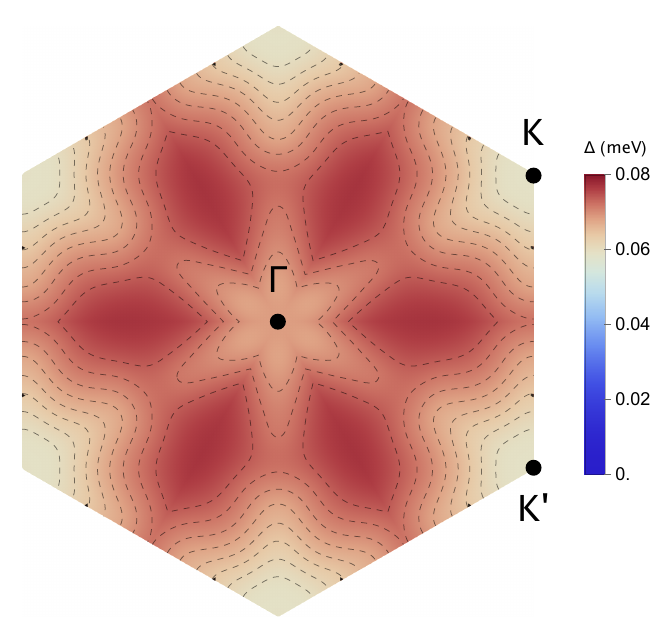}
    \caption{The momentum-dependent superconducting gap of twisted bilayer graphene for filling $n_0 = +2$  and twist angle $\theta = 1.00 \degree$.}
    \label{figure:BilayerGaps}
\end{figure}

An initial ECCSD simulation showed that the addition of correlation effects via doubles yields no difference to the Fermi level. Hence, it can be concluded to have no impact on the excited state filling and, hence, electrical conductivity, displayed in Fig. ~\ref{figure:BilayerBands}. The maximal superconducting gap was found at a twist angle of $\theta_c = 1.00 \degree$. The momentum-dependent superconducting gap at this angle and filling $n_0 = +2$ is shown in Fig. ~\ref{figure:BilayerGaps}. As previously discussed, the current approach makes no assumption about the nature of this gap, but the results point at a, roughly equal, combination of s-wave and f-wave superconductivity, which seems to challenge previous studies \cite{Lake2022, Oh2021}.

\begin{figure}
    \centering
    \includegraphics[width=0.5\textwidth]{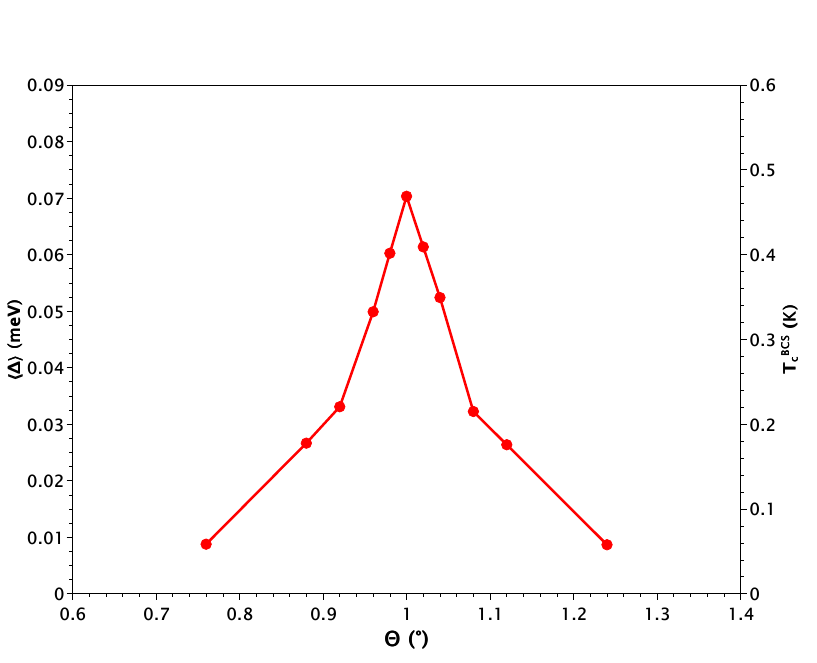}
    \caption{The mean superconducting gap over the first Brillouin zone (left scale) and the BCS estimate of the critical temperature (right scale) as a function of the twist angle of twisted bilayer graphene for a filling $n_0 = +2$.}
    \label{figure:BilayerTemperatures}
\end{figure}

Furthermore, since the superconducting gap is rather uniform with respect to the momentum, it doesn't seem unrealistic to use Bardeen-Cooper-Schrieffer (BCS) theory to estimate the critical temperature. The zero-temperature gap is related to the critical temperature as $\langle \Delta \rangle \approx 1.78 k_B T_\text{c}^\text{BCS}$. The maximal gap and, thus, the highest critical temperature is obtained at an angle $\theta_c = 1.00 \degree$ with a value of $T_\text{c}^\text{BCS} = 0.5K$. The dependence of the average gap energy on the twist angle is displayed in Fig. ~\ref{figure:BilayerTemperatures} above, along with the respective critical temperatures.

Looking at the form of the ECCSD equations, it can be seen that a non-zero superconductive gap is mathematically not possible via mean-field theory - only by accounting correlation effects via doubles and higher truncation levels - since no singles amplitudes are present.It was found that an increase or decrease of the filling from $n_0 = +2$ to either $n_0 = +1, +3$ yields an increase or decrease of the superconducting gap by $\sim 8 \%$, respectively.

The values obtained above are qualitatively very close to the experimental ones \cite{Cao2018UnconventionalS, Oh2021}, $\theta \sim 1.1 \degree$ and $T_c \sim 1K$. We have not studied the single particle spectrum, and can not make direct statements about insulating phases. However, we have not included interactions mediated by phonons. So even though inclusion of more terms in the BM model, SVD decomposition and ECC truncation would yield a more quantitative description, which would come at a very significant computational cost, one would have to include phonon-mediated interactions as well.

\section{Discussion}
\label{section:Conclusion}

It has been shown that one can study correlation effects in TBG using the coupled cluster method. This was done within the BM model with an added double metallic gate potential to account for electrostatic interactions in an experimental setting, assuming TBG is sandwiched between hBN layers. Furthermore, equations were derived for the ECCSD method, suitable for study of phase transitions in strongly correlated systems. The spectrum of monolayer graphene is successfully described to gain confidence in the method and its implementation.

For TBG, simulations at singles truncation agree with previous Hartree-Fock studies in that electrostatic effects are the leading contribution to the conductivity, because of the significant narrowing of the gaps in the band structure and impact on the Fermi level. Correlation effects at doubles truncation show insignificant contributions to the band structure and the Fermi level.

By the use of tensor decomposition and SVD, it was made possible to qualitatively simulate the superconductive gap energies and hence estimate the critical temperature and angle at which the electrical resistivity disappears. The obtained values of $\theta_c = 1.00 \degree$ and $T_c = 0.5$K matched the experimental values closely at a qualitative level. Additionally, a roughly equal combination of s-wave and f-wave symmetry was found to be present in TBG, challenging some suggestions in past research. Hence, this study presents a novel contender for the mechanism behind superconductive phases in TBG. We have found no evidence for insulating phases.

\section*{Acknowledgments}

The authors acknowledge access to computing resources provided by the Imperial College Research Computing Service \cite{ImperialHPC}.

\section*{Author Contributions}

NRW designed, supervised the project and contributed to the research. IV performed most of the coding and produced the results reported. Both authors contributed to writing the paper.

\section*{Funding Statement}

IV is funded by a Royal Society doctoral studentship from grant URF-R1-191292. NRW is funded by the Science and Technology Facilities Council under grant ST/V001116/1. 

\section*{Conflict of Interest}

The author(s) declare no conflict of interest.

\bibliography{sources}

\end{document}